\documentclass[12pt]{article}

\usepackage{sbc-template}
\usepackage[T1]{fontenc}
\usepackage[utf8]{inputenc} 
\usepackage{graphicx}
\usepackage{url}
\usepackage{natbib}
\usepackage{hyperref}

\newcommand{\titulo}{Characterizing   Synchronous  Writes   in  Stable
  Memory Devices}

\hypersetup{
     pdftitle={\titulo}, 
     pdfauthor={William B. Mingardi, Gustavo M. D. Vieira},
     pdfdisplaydoctitle=true,
     hidelinks
}

\bibpunct{[}{]}{,}{a}{}{,}
\renewcommand{\cite}[2][]{\citep[#1]{#2}}

\title{\titulo}

\author{William   B.     Mingardi\inst{1}   \and   Gustavo    M.    D.
  Vieira\inst{1}}

\address{DComp -- CCGT -- UFSCar\\
         Sorocaba, São Paulo, Brasil
         \email{williammingardi@gmail.com, gdvieira@ufscar.br}}

\begin{document}

\maketitle

\begin{abstract}
  Distributed algorithms that operate  in the fail-recovery model rely
  on   the  state   stored   in  stable   memory   to  guarantee   the
  irreversibility of operations even in  the presence of failures. The
  performance of these  algorithms lean heavily on  the performance of
  stable  memory.    Current  storage  technologies  have   a  defined
  performance  profile: data  is  accessed in  blocks  of hundreds  or
  thousands of bytes,  random access to these blocks  is expensive and
  sequential access  is somewhat better.  File  system implementations
  hide some of  the performance limitations of  the underlying storage
  devices   using   buffers   and  caches.    However,   fail-recovery
  distributed algorithms  bypass some of these  techniques and perform
  synchronous writes to be able to tolerate a failure during the write
  itself. Assuming the  distributed system designer is  able to buffer
  the  algorithm's  writes,  we  ask   how  buffer  size  and  latency
  complement  each other.   In  this  paper we  start  to answer  this
  question by characterizing the  performance (throughput and latency)
  of  typical stable  memory  devices using  a  representative set  of
  current file systems.
\end{abstract}

\section{Introduction}

Paxos~\cite{lamport98,   lamport06a}   is  a   distributed   consensus
algorithm for  asynchronous distributed  systems.  It  can be  used to
create replicas of distributed services to ensure greater availability
of the service as a whole~\cite{schneider90}. This algorithm and other
similar  consensus algorithms  are  used in  mission critical  systems
around the world~\cite{chandra07, hunt10}.

The  Paxos  algorithm  was  designed to  tolerate  faults  assuming  a
\emph{fail-recovery} fault model, in  which the processes that compose
the  distributed system  can fail  by crashing  and later  recover and
return to operation~\cite{cachin11}. When one of these failures occurs
the  process  stops operating  and  loses  all  state stored  in  main
memory. It can later return to  normal operation, but can only rely on
state stored  in stable memory, such  as the hard disk.   This failure
model is very  interesting because it is a  suitable representation of
the way computers crash, are restarted and reload the operating system
and running applications.

Paxos   and  similar   consensus  algorithms   that  operate   in  the
fail-recovery  model rely  on the  state  stored in  stable memory  to
guarantee the  irreversibility of operations  even in the  presence of
failures. This irreversibility is crucial for the consistency of these
distributed algorithms~\cite{lamport06a}.  However, accesses to stable
memory  are usually  much slower  than  accesses to  main memory,  and
algorithms in the fail-recovery model  minimize the use of this costly
resource. For  example, Paxos has two  writes to stable memory  in its
critical path~\cite{lamport98}.  This  way, performance of distributed
algorithms  in  the  fail-recovery  model will  lean  heavily  on  the
performance of stable memory.

Currently there are two main  implementations of stable memory in use:
spinning  disks and  flash  memory.  Both  technologies are  radically
different, but have a similar performance profile: data is accessed in
blocks  of hundreds  or thousands  of  bytes, random  access to  these
blocks   is  expensive   and  sequential   access  is   somewhat  more
efficient~\cite{ruemmler94,min12,chen09}.  Access to  these devices is
mediated   by   the   operating   system   using   the   file   system
abstraction. File system implementations  take the performance profile
of storage  devices into  account, and  are able to  hide some  of the
performance  limitations of  the underlying  device using  buffers and
caches.   However, fail-recovery  distributed algorithms  require that
the data  being written is committed  to disk before proceeding  to be
able to tolerate a failure during the write itself.  As a consequence,
every write must be a  \emph{synchronous write}, usually achieved by a
call to \textsl{fsync} or similar  system call.  Issuing a synchronous
write,  however,  conflicts  with  most strategies  employed  by  file
systems to hide the  performance cost of stable memory~\cite{jannen15,
  yeon18}.

Considering  the limitations  of  file  systems, we  ask  how can  the
designer  of a  distributed  system maximize  the  performance of  the
available storage devices.   To maximize something it  is necessary to
first   establish   appropriate   metrics.   If   we   only   consider
\emph{throughput} of  the storage device, its  physical implementation
and the way file systems are  implemented suggest that we should write
as much data  as possible in any single synchronous  write.  This way,
the  file  system will  be  able  to  fill  its buffers,  to  allocate
relatively  large   extents  of  contiguous  blocks   and  the  device
controller  will have  enough data  to optimize  the execution  of the
write. However, from the point of view of the distributed algorithm, a
big  write will  mean a  larger  \emph{latency} for  the operation  to
complete.  Making the question much  more interesting is the fact that
due to  the way  file systems are  implemented, this  latency increase
\emph{is  not}  directly proportional  to  the  amount of  data  being
written. Specifically,  synchronous writes  of only  a few  bytes each
will reduce  the throughput of  the disk  to almost zero,  while large
writes will hardly affect the latency of the operation.

Thus, if the distributed system  designer is able to \emph{buffer} the
algorithm's writes  to stable memory~\cite{vieira10} and  wants to use
storage devices at optimum capacity, we need to understand exactly how
buffer size and latency complement each other.  To this end, we should
evaluate the full programming environment, including operating system,
file  system  and  storage  device.   In  this  paper  we  start  this
investigation  by  characterizing   the  performance  (throughput  and
latency) of typical stable memory devices, a spinning disk and a solid
state drive, using a representative set of current file systems.  Many
works     evaluate     file      system     and     storage     device
performance~\cite{chen09,   min12,   jannen15,  sweeney96,   mathur07,
  rodeh13, lee15}, however these works  usually assume the file system
will be able to handle any type of operation the application requires.
This  work is  different  in the  sense that  we  investigate how  the
distributed  system designer  can  help the  file  system and  storage
device  better   handle  the   use  pattern  created   by  distributed
algorithms.

This paper is organized as follows. In Section~\ref{sec:background} we
describe  the basic  assumptions  of our  research  and related  work.
Section~\ref{sec:experiments}  describes  our experimental  setup  and
Section~\ref{sec:results} shows the experimental data and analysis. We
present some concluding remarks in Section~\ref{sec:conclusion}.

\section{Background}
\label{sec:background}

\subsection{Synchronous Writes}

Storage  devices are  orders  of magnitude  slower  than main  memory.
Random access of data at the byte level is prohibitively expensive and
the best  performance of  these devices can  be achieved  by accessing
data   in   sequential   blocks    of   hundreds   or   thousands   of
bytes~\cite{ruemmler94,min12,chen09}.   This  reality has  shaped  the
design of  file systems  for decades.  Among  the techniques  used are
caching of recently accessed blocks in memory for reading and writing,
and the pre-fetching  of blocks that are likely to  be accessed in the
future~\cite{rosenblum92}.   Write  caches   are  particularly  useful
because they allow successive small writes to happen to the same block
while requiring only one single write to the storage device.

The  negative consequence  of  the use  of write  caches  is that  the
application can't  be sure if the  data it has written  is effectively
recorded in the storage device. If a failure occurs after the write is
issued but before the data is  written to stable memory this write may
be  lost.  For  many applications  this behavior  is acceptable,  as a
crash means the application has already unexpectedly stopped and it is
inconsequential if  this failure happened  before or after  the write.
After  a   potential  recovery,  that  application   just  redoes  the
interrupted  write.  However,  some  applications  that interact  with
external systems have the requirement that  a write they had made must
not  be forgotten.   One simple  example of  these applications  is an
automated  teller machine  that can't  undo  the fact  it has  already
dispensed  money.  Distributed  algorithms have  similar requirements,
but instead of  money they can't take back messages  that were already
sent~\cite{lamport06a}.

Operating  systems support  the  implementation  of applications  that
require  that  a  write  be  committed to  stable  storage  through  a
\textsl{fsync} or similar  system call. This call will  force the file
system  to  write to  disk  all  buffers  it  keeps in  memory.   Very
frequently  metadata  and  caches  are also  flushed  and  immediately
written  to  the   underlying  device~\cite{jannen15,  yeon18}.   This
satisfies the requirement of the  application, but have a considerable
impact  on  the  performance  of   the  file  system.   Moreover,  the
application  itself can  be victim  of  its own  access patterns.   To
explain  how  this   happens,  let's  consider  the   situation  of  a
distributed algorithm that  needs to write $x$ bytes  for each message
it  sends, where  $x$ is  small  number.  If  this application  writes
synchronously these few bytes, it will have to wait for the write of a
complete file system  block plus any metadata changes  required by the
file system.  If we assume the latency of writing a block is $l_b$ and
the latency of metadata update is  $l_m$, the throughput of this write
is:
\[\frac{x}{l_b + l_m}\]

However, if  the application is able  and decides to batch  a group of
messages sending  them at the  same time, it  can make a  single write
corresponding to  the record  of all  messages sent  at once.  If this
group has $y$  messages and the size of bytes  written $xy$ is smaller
than the file system block size, we can assume the latency $l_b + l_m$
will remain  constant. Under this  assumption, the throughput  of this
batched write will be:
\[\frac{xy}{l_b + l_m}\]
that is $y$ times larger than  the throughput of recording the send of
a single message.  Depending on the  size of the file system block and
the size of this \emph{application buffer} of $xy$ bytes, the increase
of throughput obtained can be orders of magnitude larger than the case
where $y =  1$.  Also, the latency  $l_b + l_m$ can  be considered the
minimum  latency required  to  perform a  synchronous  write.  We  can
improve  the throughput,  but  the  best we  can  achieve under  ideal
conditions is to maintain this minimum latency.

The  assumption  that $l_b  +  l_m$  will  remain  constant is  a  bit
optimistic, though.   At the least,  as $xy$ increases  eventually the
size  of   the  application  buffer   will  exceed  the   file  system
block. Moreover, the amount of metadata updates required will increase
as  the  number of  file  system  blocks  touched  by the  write  also
increases. Then, as we increase  $y$ the latency will eventually creep
up, matching some  of the throughput gains we  obtained.  The question
an application programmer has to face  now is how many messages should
one batch in an application buffer  to increase throughput to an ideal
level while not increasing latency disproportionately.

This trade-off of throughput and latency is not new, actually it is at
the     heart     of     many    computing     devices     we     rely
on~\cite{patterson04}. What  is interesting in  this case is  that, at
least for  implementations of  distributed algorithms and  the current
designs for file  systems, this trade-off falls squarely  in the hands
of the application programmer.  Two techniques that have been employed
at the \emph{application level} is to use a log-like structure to keep
writes   sequential  and   accumulating   data  in   a  large   enough
buffer~\cite{vieira10}.        These       are      common       sense
approaches~\cite{patterson04},  but that  lack  a  solid framework  to
evaluate their effectiveness.

\subsection{Related Work}

Management of  synchronous writes is  a problem usually faced  by file
system       designers       with      respect       to       metadata
consistency~\cite{rosenblum92}.   Coalescing write  operations in  the
buffer cache is usually done internally by the file system, instead of
the application.  For this reason, studies considering the performance
of synchronous writes under  different \emph{application} buffer sizes
are practically nonexistent.

From the point of view of  file system design and implementation, many
studies focus on making metadata  updates more efficient. One approach
is   a   log-structured   file   system~\cite{rosenblum92}   such   as
F2FS~\cite{lee15}. This  file system  tries to transform  random write
patterns  in  sequential  ones  by  avoiding  to  change  metadata  in
place. Instead,  changes are written  in a sequential log  of changes,
properly  indexed   for  later  access.   F2FS   designers  show  that
synchronous writes  make or break  a file system performance  and that
coalescing writes can be very efficient~\cite{lee15}.  F2FS writes data
in  chunks   of  512   kB  while   a  recent   file  system   such  as
ext4~\cite{mathur07} usually  writes 4  kB. F2FS  is a  very efficient
file system, however there  is a limit to what the  file system can do
alone. Many small synchronous writes  will kill the performance of any
file system, including F2FS, as we show in Section~\ref{sec:results}.

Research  proposing  new  file   systems  usually  is  accompanied  by
performance  evaluations~\cite{sweeney96,  mathur07, rodeh13,  lee15},
but the  focus are workloads  that do not include  synchronous writes.
One  exception  is~\cite{lee15}  that  shows  performance  data  of  a
workload with \textsl{fsync} on top of F2FS.  However, this evaluation
assumes many  concurrent threads  providing enough  data to  batch the
writes in  larger buffers.   We are interested  in the  performance of
single-threaded synchronous  writes that  reflect more  accurately the
latency expected by a distributed  algorithm.  Other examples of works
that perform extensive performance evaluations of storage systems, but
do   not   consider   synchronous  writes,   are   \cite{chen09}   and
\cite{jannen15}.

Write-optimized file  systems such as BetrFS~\cite{jannen15}  could be
used  to offer  efficient  writes to  the  application, regardless  of
buffer  size.  However,  these  file systems  still  suffer from  many
limitations~\cite{jannen15}  and aren't  yet in  widespread use.   The
\textsl{fsync}  optimizations  present  in  \cite{yeon18}  would  help
reduce the inherent latency of  performing a \textsl{fsync}. This work
complements   our  results   in   the  sense   that  improvements   in
\textsl{fsync}  latency  will  be   a  constant  factor  reduction  of
$l_b + l_m$, but that alone won't change the problem of low throughput
with small buffers.

\section{Experimental Characterization}
\label{sec:experiments}

\subsection{Overview}

We  want   to  characterize  the  performance   of  secondary  memory,
specifically  we  are interested  in  the  performance of  sequential,
synchronous writes  that ignore  the file  system buffers.  The target
application is  a distributed  algorithm, such as  Paxos, that  has to
perform synchronous writes (\textsl{fsync}) to stable memory before it
can proceed  with its execution.   We also assume that  this algorithm
processes many  requests in parallel  and is  thus able to  assist the
file system by coalescing its own writes in larger batches at the cost
of increased latency for its operations.

The  focus of  the characterization  is the  trade off  between stable
memory throughput and latency as the  application sets the size of its
write buffer:  the larger the  buffer, the larger both  throughput and
latency. The objective is to give the application designer a tool that
can  be  used  to  discover,   for  a  combination  of  stable  device
implementation  and file  system,  which is  the  optimal buffer  size
considering the latency requirements of the application.

We have performed tests with application  buffers ranging from 4 kB to
16 MB,  as these sizes cover  the point where latencies  start to grow
proportionally to the buffer size.  The tests write sequentially a new
file of 16 MB and we have observed that bigger files do not change the
throughput and  latency data  observed.  The  experiments were  run on
different combinations of stable  memory technology (spinning disk and
flash memory) and files systems (ext4, XFS, BTRFS, F2FS).

\subsection{Iozone}

The  tests  were  run  using the  Iozone  multi-platform  file  system
benchmark  tool.\footnote{\url{http://www.iozone.org/}} This  tool was
selected because it is portable to  many operating systems, it can run
many  different  tests  including   reading,  writing,  rereading  and
rewriting, besides being highly  customizable.  We configured the tool
to  perform a  sequential synchronous  write test  with the  following
parameters:

\begin{description}
\item[Test type:] The test is a sequential write of a file (-i 0).
\item[Buffer size:] The  buffer sizes tested range from 4  kB to 16 MB
  (-a).
\item[File size:] The file written has 16 MB (-n 16m -g 16m).
\item[Synchronous write:]  The writes  bypass the file  system buffers
  and go directly  to disk and the wait time  of the synchronous write
  is added to the operation latency (-o -I -e).
\item[Latency tests:] The default for Iozone is to measure throughput,
  but we also tested the latency of operations (-N).
\end{description}

Unfortunately, the Iozone documentation is not clear about how many
iterations it performs when calculating the throughput and latency for
each block size.  More importantly, we  did not have access to the raw
data or  even indirect measures  beyond the average, such  as standard
deviation.  To achieve more rigorous  results we decided to run Iozone
repeatedly (30 times)  and treat the output of each  run as a separate
data point. We then calculated the average and standard deviation from
this data. Our  approach was validated when we observed  that for some
experimental parameters the variation of the readings for many runs of
Iozone was considerable (Section~\ref{sec:results}).

To automate the process of repeatedly running Iozone we created a Bash
script that runs  the benchmark, collects the data  and aggregates the
result, calculating the average  and standard deviation.  Other aspect
automated by the  script is that Iozone  has to be run  twice for each
data point: the  first for throughput and the second  for latency. For
each run the raw  output of Iozone is parsed by Awk  and a simple text
file describing  this run is  created, containing for each  block size
the  relevant metric  (throughput  or latency).   After  all runs  are
performed, these files  are coalesced in a single  file containing the
average  and   standard  deviation   for  each   block  size   of  the
measurement. Finally, all intermediary data  and the final results are
archived in  a single compacted  file to allow future  reprocessing of
the data, if necessary.

This way,  the script  created integrates all  steps to  reproduce our
data, and  more importantly, is  the seed  for an automated  tool that
could be used  to inform an application programmer  about the behavior
of the file system and storage device being used.

\subsection{Experimental Setup}

The tests were run in a desktop computer running 64 bits Linux (Fedora
25), using version 4.12.9 of the Linux kernel. All tested file systems
were  at the  standard versions  found  in this  kernel release.   The
computer had  an Intel Core 2  6300 processor, with a  1.86 GHz clock,
and 4 GiB of RAM. For storage devices to be tested we installed in the
computer two units that throughout the text are going to be referenced
as HDD and SSD:
\begin{description}
\item[HDD:] HITACHI HDS72101 7200 rpm hard disk (1 TB),
\item[SSD:] Samsung 850 solid state drive (256 GB).
\end{description}
Both drives were  connected to a SATA II interface  and the tests were
run in  a dedicated 19  GiB partition in  each drive. The  drives were
both dedicated to the test, there was  no system files or user data in
them and no competing accesses.

To validate  that our experimental setup  and data were robust  and to
assess the  portability of the test  script, we also run  tests with a
different type of storage device  in another machine. These tests were
run in  a Vostro  5740 laptop  running 64  bits Linux  (Ubuntu 16.04),
using version  4.4.0 of the  Linux kernel.   This laptop had  an Intel
Core i5-4210U processor, with  a 1.70 GHz clock, and 4  GiB of RAM. In
this machine we ran tests on a  SD card attached to an integrated card
reader, referenced in the text as SDCARD:
\begin{description}
\item[SDCARD:] SandDisk Ultra SD card (32 GB).
\end{description}
In  this  device   the  tests  were  run  in  a   dedicated  1.44  GiB
partition. Although this device is not representative of the type of
devices usually  employed to support distributed  algorithms, we found
that the data  obtained helps to support some observations  we made as
more general and applicable to a wide range of devices.

Even though our  test load uses synchronous writes to  ensure data was
stable in  the drive before  the application  can continue, we  do not
bypass  the file  system  and  go direct  to  the  device.  Thus,  the
performance of the file system is also a significant factor in overall
performance of the target applications.  To assess how big this factor
is,  we  selected  a  representative  set  of  current  file  systems:
XFS~\cite{sweeney96},  ext4~\cite{mathur07}, BTRFS~\cite{rodeh13}  and
F2FS~\cite{lee15}. XFS and ext4  represent modern implementations of a
``classic'' file system, while BTRFS is  a more recent design. F2FS is
a log-structured file system tailored for flash-based devices.

\section{Results}
\label{sec:results}

In this section  we first present the throughput and  latency data for
each of the  devices and file systems tested. Using  this data we show
how to  choose a  buffer size appropriate  for a  specific distributed
algorithm.   All charts  in this  section have  logarithmic-scaled $x$
axis, because data on buffer size  was measured by doubling the buffer
size. This  way we can  cover a larger range  of buffer sizes  in less
time, but this distorts the data. To compensate for this, the $y$ axis
is also in  logarithmic scale. In our discussion of  the results, when
comparing two  file systems we use  an independent t-test to  test for
statistic significance, with $p < 0.001$ as threshold.

\subsection{Hard Disk}

\begin{figure}[htb]
  \centering
  \includegraphics[width=0.8\textwidth]{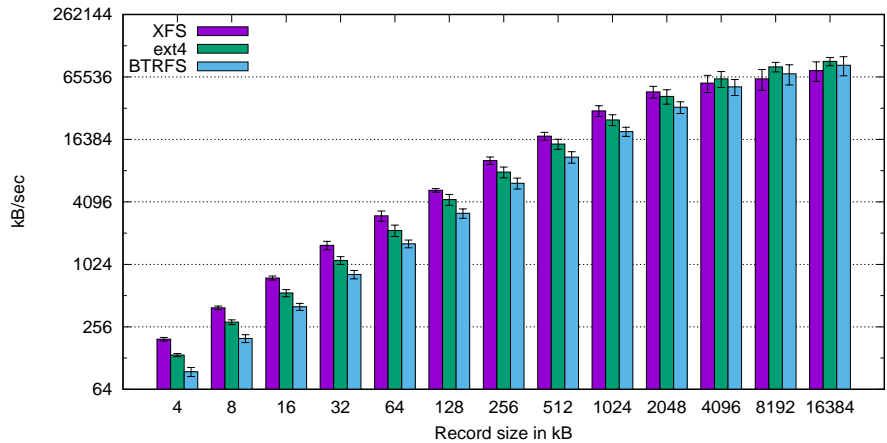}
  \caption{HDD Throughput}
  \label{fig:hd_ext4_btrfs_xfs}
\end{figure}

\begin{figure}[htb]
  \centering
  \includegraphics[width=0.8\textwidth]{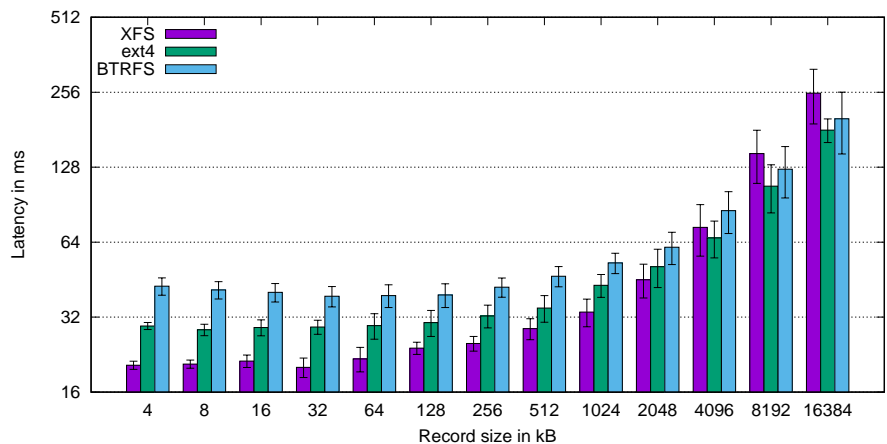}
  \caption{HDD Latency}
  \label{fig:hd_ext4_btrfs_xfs_op}
\end{figure}

The    results     for    the     HDD    device    are     shown    in
Figure~\ref{fig:hd_ext4_btrfs_xfs}           (throughput)          and
Figure~\ref{fig:hd_ext4_btrfs_xfs_op}  (latency),  comparing the  XFS,
ext4 and  BTRFS file systems.   The first  observation we can  make is
that, as expected, throughput  increases proportionally to buffer size
while latency remains mostly constant for small buffer sizes.  In this
range, performance is dictated by synchronous write performance of the
device  and  efficiency of  the  file  system.  As  buffers  increase,
however, latencies start to raise  as memory and disk throughput start
having  a  larger impact  on  the  performance.  For  larger  buffers,
throughput remains constant while  latency increases proportionally to
buffer    size.    This    behavior   is    also   expected,    as   a
throughput-saturated  disk  will take  more  time  to write  a  larger
buffer.

The  behavior  considering  very  small  and  very  large  buffers  is
predictable, while the transition between the  two is much less so. In
Section~\ref{sec:ratio} we will discuss this transition in more depth,
but first we are going to make some observations about the performance
of the file systems in the HD device. For small buffers, BTRFS has the
worst  latency, XFS  has the  best, at  about half  of the  latency of
BTRFS,  and ext4  stays in  the  middle.  This  latency dominates  the
throughput and  the relative performance  of these file  systems stays
roughly the same for small buffers:  XFS is the best, followed by ext4
and BTRFS. We  don't have data to support this,  but we speculate this
difference in  performance is  due to the  management of  metadata for
each file system, with BTRFS showing the largest overhead.

For larger  buffers we are  approaching the maximum throughput  of the
underlying device. With a saturated disk file system differences start
to disappear  and optimizations to  writing of bulk data  and metadata
start to  make a  difference. This  appears to  be specially  true for
BTRFS,    which    takes    advantage   of    larger    file    system
blocks~\cite{rodeh13}.  There are still some noticeable differences in
the  averages, but  the  standard deviation  increased  as well.   For
instance, for the buffer size of 4096  kB that marks a point where the
average  throughput of  ext4  surpasses the  throughput  of XFS,  this
difference is not statistically significant
($t_{(30)} = 2.02, p = 0.049$).

\subsection{SSD}

\begin{figure}[htb]
  \centering
  \includegraphics[width=0.8\textwidth]{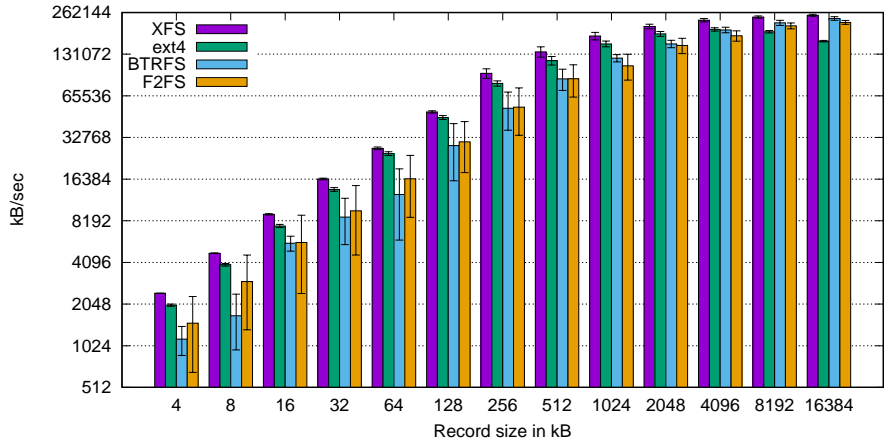}
  \caption{SSD Throughput}
  \label{fig:ssd_ext4_btrfs_xfs_f2fs}
\end{figure}

\begin{figure}[htb]
  \centering
  \includegraphics[width=0.8\textwidth]{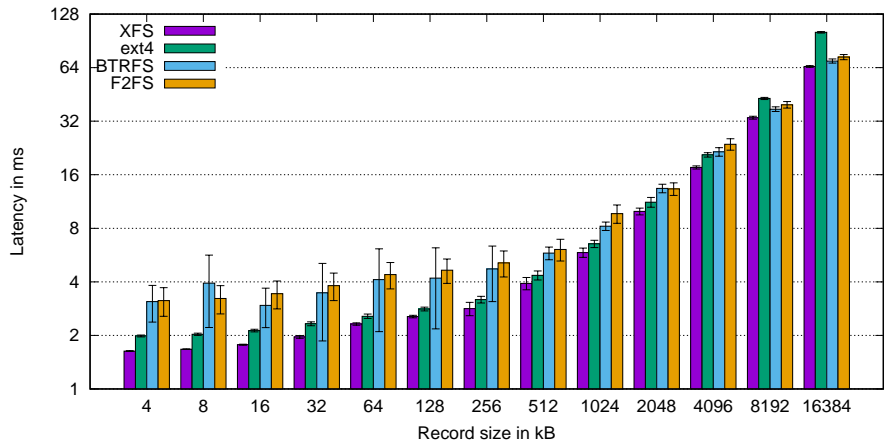}
  \caption{SSD Latency}
  \label{fig:ssd_ext4_btrfs_xfs_f2fs_operations}
\end{figure}

The    results     for    the     SSD    device    are     shown    in
Figure~\ref{fig:ssd_ext4_btrfs_xfs_f2fs}        (throughput)       and
Figure~\ref{fig:ssd_ext4_btrfs_xfs_f2fs_operations}         (latency),
comparing  the XFS,  ext4,  BTRFS  and F2FS  file  systems.  The  data
confirms    that,   despite    their   considerable    difference   in
implementation,  both HDD  and SSD  share a  very similar  performance
profile.  We  can observe increasing throughput  with constant latency
for smaller  buffers, and constant throughput  with increasing latency
for larger buffers.  What takes HDD  and SSD apart is the magnitude of
the performance, with  throughput and latency of SSD  about 4x better.
Moreover, the SSD  device reaches its maximum  throughput with smaller
buffers than the HDD. This means that  this class of device has a real
advantage  for use  with  distributed algorithms  because the  maximum
throughput  can be  achieved  with  lower latency,  as  we discuss  in
Section~\ref{sec:ratio}.

Regarding  the relative  performance of  the tested  file systems,  we
observed in the SSD device a  similar pattern found in the HDD device.
For small buffer sizes, XFS and  ext4 have a clear advantage, with XFS
leading by a small margin. For example, with buffer size of 128 kB XFS
achieves 50127  kB/s while  ext4 manages  45675 kB/s,  a statistically
significant difference ($t_{(30)} = 13.8, p < 0.001$).  Both BTRFS and
F2FS  have  weak  performance  with small  buffers,  probably  due  to
metadata overhead.  F2FS in  particular buffers concurrent synchronous
writes and is not particularly suited for the single-threaded workload
we tested~\cite{lee15}.   Moreover, BTRFS  and F2FS  show a  very high
standard deviation, probably indicating that metadata overhead is not
constant and probably has some infrequent high-cost operations.

As observed for  the HDD, as we  increase the size of  the buffers for
the SSD,  file systems differences tend  to get smaller.  But,  as the
SSD device shows  a smaller standard deviation,  these differences are
more  consistent.  For  example,  for  buffer  size  of  4096  kB  the
throughput  of XFS  is 230612  kB/s while  the throughput  of ext4  is
197912     kB/s,     a    statistically     significant     difference
($t_{(30)} = 20.8,  p < 0.001$).  One unexpected  observation was that
ext4  had a  noticeable  and  consistent drop  in  throughput for  the
largest buffers.  We  don't have data to support it,  but we speculate
as the throughput increases ext4 uses more CPU and may start to be CPU
bound.

\subsection{SDCARD}

\begin{figure}[htb]
  \centering
  \includegraphics[width=0.8\textwidth]{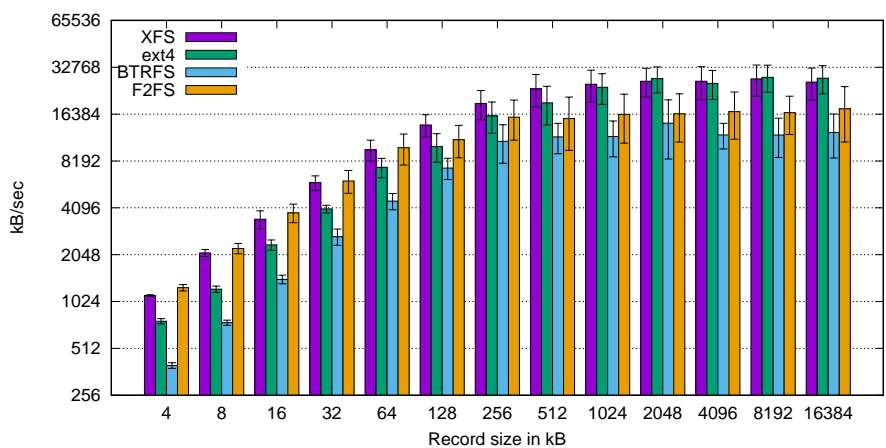}
  \caption{SDCARD Throughput}
  \label{fig:sdcard_ext4_btrfs_xfs_f2fs}
\end{figure}

\begin{figure}[htb]
  \centering
  \includegraphics[width=0.8\textwidth]{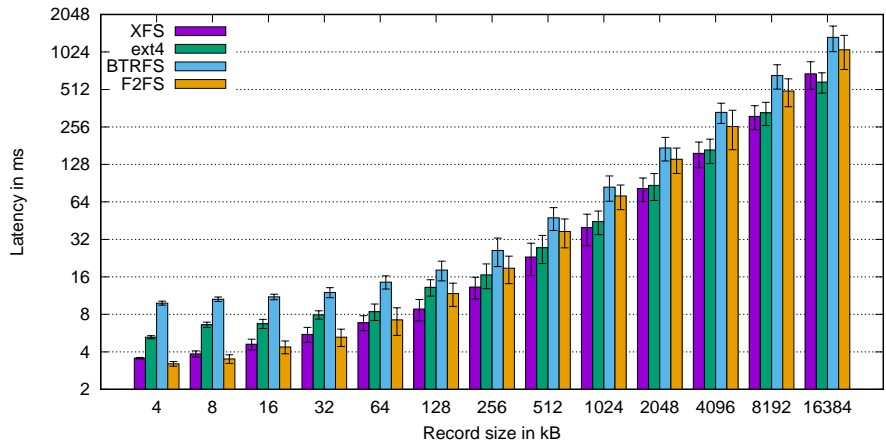}
  \caption{SDCARD Latency}
  \label{fig:sdcard_ext4_btrfs_xfs_f2fs_operations}
\end{figure}

The    results    for    the    SDCARD    device    are    shown    in
Figure~\ref{fig:sdcard_ext4_btrfs_xfs_f2fs}      (throughput)      and
Figure~\ref{fig:sdcard_ext4_btrfs_xfs_f2fs_operations}      (latency),
comparing  the XFS,  ext4, BTRFS  and F2FS  file systems.   The SDCARD
device  is  a  low  capacity,  removable  device,  with  a  basic  FTL
(\emph{flash translation  layer}~\cite{chen09,min12}) and consequently
lower throughput  and higher  latency than the  SSD device.   The data
confirms a  low maximum throughput,  but the  device is able  to reach
this  throughput  with smaller  buffers.   As  a consequence,  at  the
maximum throughput the latency is surprisingly low.

F2FS has specially low latencies up  to 32 kB buffers and consequently
a higher throughput, beating XFS by a small margin. This difference is
only    statistically    significant    up    to    8    kB    buffers
($t_{(30)} = 3.9, p = 0.0002$).  Nonetheless, this shows the F2FS file
system is particularly optimized for this class of storage device.

For larger  buffer sizes, F2FS  disappointingly plateaus at  about two
thirds of  the final  throughput achieved  by both  XFS and  ext4. For
these   larger  buffer   sizes   the   standard  deviation   increases
considerably and performance differences between  XFS and ext4 are not
statistically  significant. With  a buffer  size of  1 MB  XFS reaches
25469 kB/s  while ext4 throughput  is 24368 kB/s, a  not statistically
significant difference ($t_{(30)} = 0.75, p = 0.454$).  XFS is however
better than  F2FS for larger buffers.   For the same 1  MB buffer F2FS
throughput  is  16321  kB/s,  and  XFS  has  a  statistic  significant
advantage ($t_{(30)} = 6.12, p < 0.001$).

\subsection{Balancing Throughput and Latency}
\label{sec:ratio}

With the throughput  and latency curves obtained with  our benchmark a
system designer can  pick the required buffer  size for implementation
of a  distributed algorithm.   If the application  needs to  move data
with a  minimum throughput,  the designer  should choose  the smallest
buffer  that achieves  the desired  throughput, if  possible.  If  the
application requires responses to be  sent with a maximum latency, the
designer should  choose the largest  buffer that doesn't  violate this
limit,  if  possible.   However,  matters   aren't  as  clear  if  the
application hasn't any hard limit  on throughput and latency, but only
a general desire to \emph{optimize} the balance between the two.

Each metric  we have measured  has a transition:  throughput increases
until  it  reaches  maximum  device throughput  and  then  stabilizes;
latency  is stable  until transfer  costs start  to dominate,  then it
starts to increase.   The two transition points do  not coincide, they
happen in different buffer sizes.  For example, considering XFS on the
HDD the throughput starts  to level off at buffer sizes  of 4096 kB to
8192 kB, while  latency starts picking up with buffer  sizes of 512 kB
to 1024 kB.
 
\begin{figure}[htb]
  \centering
  \includegraphics[width=0.8\textwidth]{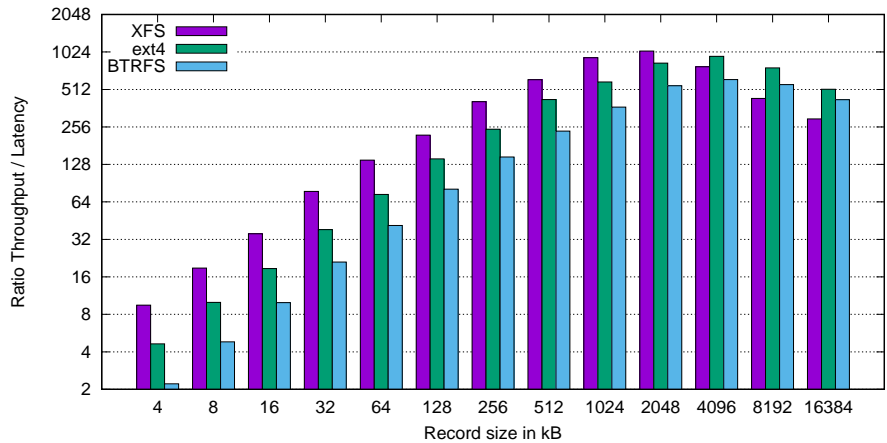}
  \caption{HDD Ratio}
  \label{fig:hd_ratio}
\end{figure}

\begin{figure}[htb]
  \centering
  \includegraphics[width=0.8\textwidth]{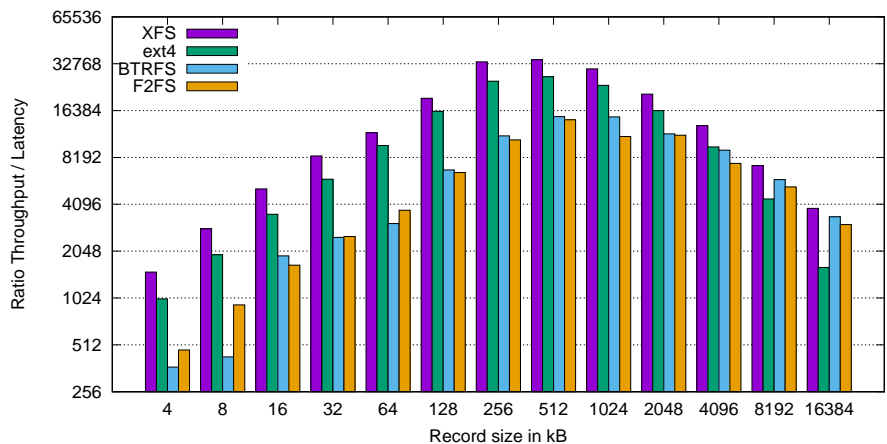}
  \caption{SSD Ratio}
  \label{fig:ssd_ratio}
\end{figure}

\begin{figure}[htb]
  \centering
  \includegraphics[width=0.8\textwidth]{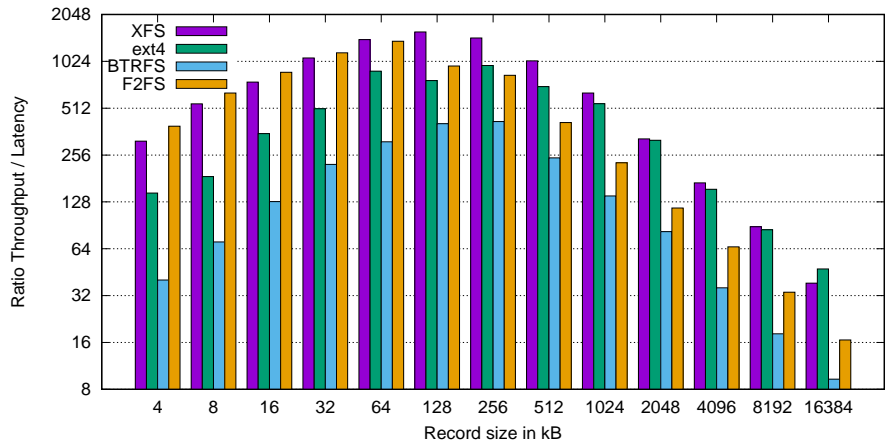}
  \caption{SDCARD Ratio}
  \label{fig:sdcard_ratio}
\end{figure}

To try and capture the tradeoff between throughput and latency in face
of the  fact that  the tipping  point of each  metric is  different we
introduce  a  combined  metric,  the  \emph{ratio}  of  throughput  by
latency. This new  metric gives a rough idea  of the \emph{efficiency}
of a buffer size, indicating how many units of throughput one can gain
for  each  unit  of latency  introduced.   Figures~\ref{fig:hd_ratio},
\ref{fig:ssd_ratio}  and \ref{fig:sdcard_ratio}  show this  new metric
for the devices and file systems tested.

In a  sense, the  curve observed  in these  figures can  be seen  as a
combination of the throughput and  latency curves. The ascending slope
represents  the  phase  which  increasing buffer  size  will  increase
throughput more than  it increases latency.  The plateau  is the phase
where latency and throughput increase in the same rate. The descending
slope represents the phase where the increase in throughput in smaller
than the increase in latency.  The maximum of the curve represents the
best throughput  for each  unit of latency.   If one  considers larger
buffers,   the  latency   is  increased   but  the   throughput  won't
improve. This is the point where throughput is maximized requiring the
least of latency.

Using this metric as guide, we can observe that the SSD device is more
capable of handling the loads of distributed algorithms because it can
saturate its  write throughput  with smaller buffers.   The SSD  has a
superior absolute performance, but  the ratio of throughput/latency is
more relevant  in this case.  As  an example of this,  take the SDCARD
device. As  a low end device,  without the sophisticated FTL  found in
the  SSD, its  performance  parameters are  arguably  inferior to  the
HDD. Maximum  throughput of the SDCARD  is about half of  the HDD, but
latency   is   about   half.     As   a   consequence,   the   optimum
throughput/latency ratio is about the  same for both devices, but with
the  SDCARD showing  latencies  that are  10x  smaller. Throughput  is
lower, however, but  only about 4x lower. Thus, for  an application in
which  this lower  throughput is  acceptable, surprisingly  the SDCARD
device would be an interesting choice.

\section{Conclusion}
\label{sec:conclusion}

Distributed  algorithms in  the  fail-recovery  failure model  require
efficient access to stable memory.  This efficiency is measured by the
trade-off between the  latency of each write to stable  memory and the
total throughput of  writes. A distributed system  programmer wants to
balance the two by correctly sizing  the buffers sent to be written by
the  file  system, ideally  achieving  a  target throughput  with  the
minimum latency possible.  To aid  in this task, we have characterized
the performance profile  of typical stable memory  devices, a spinning
disk and  a solid state drive,  using a representative set  of current
file systems.

Our data show that the performance of the studied storage devices show
three distinct phases as application  buffer size grows.  In the first
phase throughput increases while latency stays approximately constant.
In the  second phase  throughput and latency  increase proportionally.
In the  third phase  throughput reaches a  maximum and  latency starts
increasing.  In general terms, a  designer should choose a buffer size
in the first state with the minimum latency that respects the required
throughput of the application.

With  respect  to the  devices  tested,  the  ones that  saturate  the
throughput  with  smaller  buffer  sizes tend  to  offer  the  smaller
latency.  The SSD device was the  best in this respect, with very good
throughput and latency figures. Surprisingly, the SDCARD device showed
a  very interesting  balance between  throughput and  latency, despite
being a  low performance  device.  With respect  to file  systems, the
ones that handle  large buffers efficiently were  the best performers.
In particular, XFS showed a very consistent performance.

\bibliographystyle{apalike}
\bibliography{sequential.throughput.paper}

\end{document}